\documentclass[showpacs,showkeys,onecolumn,notitlepage]{revtex4-1}
\usepackage{makeidx}
\usepackage{amsmath}
\usepackage{eurosym}
\usepackage{amssymb}
\usepackage{graphicx}
\usepackage{dcolumn}
\usepackage{bm}
\usepackage{rotating}
\usepackage{natbib}

\begin{document}

\title{Interlayer Heat Transfer in Bilayer Carrier Systems}
\author{Mika Prunnila}
\email{mika.prunnila@vtt.fi}
\author{Sampo J. Laakso}
\affiliation{VTT Technical Research Centre of Finland, P.O.Box 1208, FIN-02044 VTT,
Espoo, Finland}
\date{\today }

\begin{abstract}
We study theoretically how energy and heat are transferred between the
two-dimensional layers of bilayer carrier systems due to near-field
interlayer carrier interaction. We derive general expressions for the
interlayer heat transfer and thermal conductance. Approximation formulas and
detailed calculations for semiconductor and graphene based bilayers are
presented. Our calculations for GaAs, Si and graphene bilayers show that the
interlayer heat transfer can exceed the electron-phonon heat transfer below
(system dependent) finite crossover temperature. We show that disorder
strongly enhances the interlayer heat transport and pushes the threshold
towards higher temperatures.
\end{abstract}

\maketitle

\section{Introduction}

Interlayer momentum transfer (the drag effect) has been extensively
investigated in bilayer carrier systems, where two two-dimensional (2D)
carrier gases are separated by a thin barrier. The drag effect is a
manifestation of near-field interlayer interaction and bilayer carrier
systems provide a unique laboratory for probing charge carrier interactions
and interaction driven phases (see Refs. \cite{rojo:1999,gupta:2011} for a
review). Since the pioneering experiment of electron-electron drag between
two coupled 2D electron gas (2DEG) layers in GaAs-AlGaAs heterostructure 
\cite{gramila:1991} 2D carrier bilayers have been demonstrated in variety of
semiconductor structures. Recently, the drag effect has been experimentally
investigated also in graphene bilayer, where two single layer graphene
flakes are separated by a dielectric.\cite{kim:2011}

\begin{figure}[b]
\centering\includegraphics[width=0.87\textwidth]{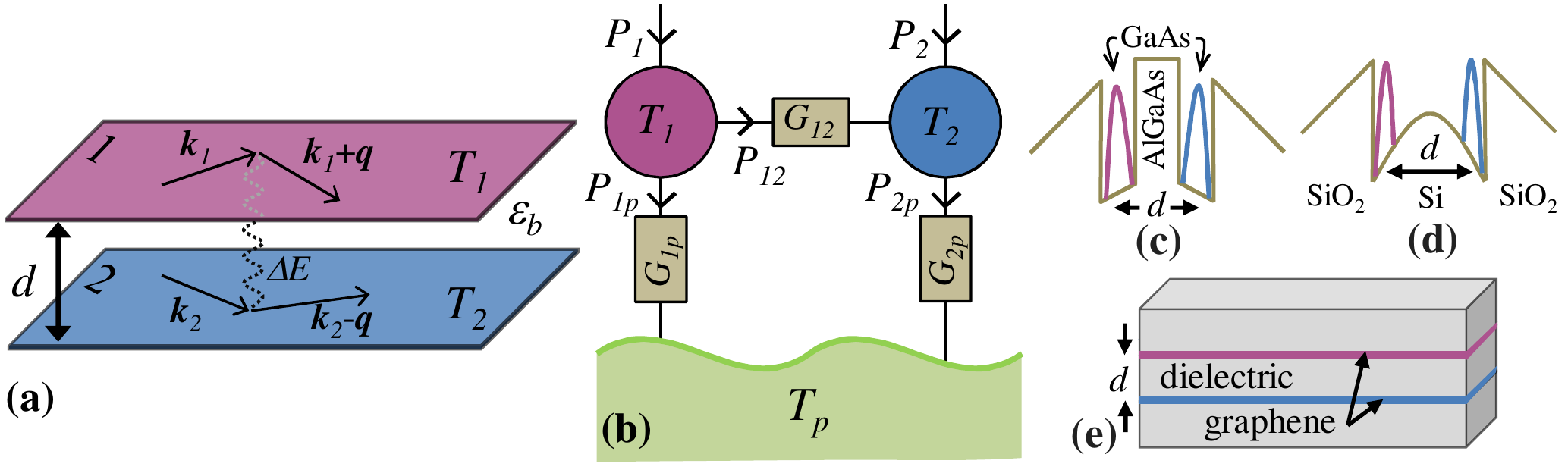}
\caption{(color online) (a) Illustration of near-field scattering processes
(momentum and energy transfer) between 2D carrier layers $1$ and $2$
separated by a distance $d$ and embedded in a solid with dielectric constant 
$\protect\varepsilon _{b}$. The layers are at local temperatures $T_{1}$ and 
$T_{2}$. A carrier in layer $1$ $(2)$ experiences momentum scattering $%
\boldsymbol{k}_{1}\rightarrow \boldsymbol{k}_{1}+\boldsymbol{q}$ ($%
\boldsymbol{k}_{2}\rightarrow \boldsymbol{k}_{2}-\boldsymbol{q}$) due to
interlayer interaction. During the process energy $\Delta E$ is transferred
between the layers. (b) The equivalent thermal circuit. $P_{L}$ is the input
heating/cooling power ($L=1,2$) and $G_{12}$ is the interlayer thermal
conductance. Due to the interlayer scattering processes $G_{12}\neq 0$ and
power (or heat) $P_{12}$ flows between the layers. Layers couple to phonon
bath, which is at temperature $T_{p}$, via electron-phonon thermal
conductance $G_{Lp}$ and power $P_{Lp}$ flows to the bath. Illustration of
conduction band diagram and electron wavefunctions of (c) GaAs and (d) Si
bilayer. (e) Graphene bilayer with dielectric barrier.}
\label{fig:schematic}
\end{figure}

The investigations of bilayer carrier systems have been focused on the drag
phenomenon, but the interlayer interaction also mediates energy and heat
transfer between the layers (see Fig. \ref{fig:schematic}) and such
near-field energy/heat transfer is the topic of the present Paper.
Considerable efforts have been devoted to understand near-field heat
transfer via different channels between bodies that are separated by a small
vacuum 
gap.\cite{pendry:1999,joulain:2005,volokitin_rev:2007} One of the most
significant heat exchange channels is built from inter-body photon coupling.
Surface excitations involving optical phonons and plasmons can play an
important role and these so-called polariton effects can strongly enhance
the near-field energy transfer.\cite{joulain:2005} Recently, a near-field
heat transfer channel arising directly from lattice vibrations has also been
proposed.\cite{prunnila:2010,altfeder:2010} Near-field heat transfer is
naturally always present between closely spaced systems, even in the case of
solid contact, but then the effect is expected to be strongly masked by
competing heat dissipation channels formed by solid heat conduction and/or
electron-phonon coupling. One of the motivation for the present work is to
challenge this line of thought and, indeed, by detailed calculations we will
show that in bilayer carrier systems the near-field heat transfer can become
the dominant interlayer heat transfer mechanism.

In this work, we derive general expression of charge fluctuation induced
interlayer energy transfer rate, which is applicable to semiconductor and
graphene bilayers. In the derivation we use perturbation theory and
fluctuation-dissipation relations. Our formula for the interlayer thermal
conductance, $G_{12}$, has strong connection to the drag resistance formulas%
\cite{flensberg:1995b}. The interlayer thermal conductance is intimately
connected to fluctuations and dissipative properties of the individual
layers. This is explicitly seen as the presence of the imaginary parts of
the layer susceptibilities in the $G_{12}$ formula and it is a manifestation
of fluctuation-dissipation theorem. Approximation formulas and detailed
calculations of $G_{12}$ in the case of screened Coulomb interlayer
interaction are presented and we show that interlayer thermal transport is
strongly enhanced due to disorder. As the layers are in the same solid there
exist competing energy relaxation channels. At the temperatures of interest
electron-phonon coupling to the bulk thermal phonons is the relevant
competing heat dissipation mechanism [see Fig. \ref{fig:schematic}(b)]. It
is shown that remarkably $G_{12}$ can dominate over the electron-phonon
coupling. Therefore, near-field heat transfer can become a dominant heat
transfer mechanism even in the case of solid contact.

\section{Theory}

In this Section we derive general expression for the interlayer thermal
conductance $G_{12}$. Then we introduce approximation formulas for $G_{12}$
and on the basis of the existing literature discuss electron-phonon
coupling, which is the competing dissipation channel.

\subsection{Interlayer thermal conductance}

The scattering events depicted in Fig. \ref{fig:schematic}(a) are mediated
by interlayer interaction which is described by matrix element $M_{%
\boldsymbol{q}}$ (to be defined later). The interlayer Hamiltonian $H$ is
given by 
\begin{subequations}
\begin{eqnarray}
H &=&\frac{1}{2A}\sum_{\boldsymbol{q}}M_{\boldsymbol{q}}\rho _{1\boldsymbol{q%
}}^{\dag }\rho _{2\boldsymbol{q}}, \\
\rho _{L\boldsymbol{q}} &=&\sum\limits_{\boldsymbol{k}}\sum\limits_{\sigma
,\sigma ^{\prime }}\sum\limits_{s,s^{\prime }}c_{\boldsymbol{k}-\boldsymbol{%
q,}s^{\prime }\sigma ^{\prime }}^{\dag }F_{\boldsymbol{k}-q,s^{\prime
}}^{\dag }F_{\boldsymbol{k},s}c_{\boldsymbol{k},s\sigma },
\end{eqnarray}%
\end{subequations}%
where $A$ is the area, $\rho _{L\boldsymbol{q}}$ is the electron density
operator for layer $L=1,2$ and $c_{\boldsymbol{k,}s\sigma }^{(\dag )}$ is
the electron annihilation (creation) operator. Variables $\boldsymbol{k},$ $%
s $ and $\sigma $ are wavevector, band index and spin index, respectively
(here we will assume spin degeneracy). All electron variables depend on the
layer index $L$, but this is typically not written explicitly (e.g. $%
\boldsymbol{k=k}_{L}$). Factor $F_{\boldsymbol{k},s}$ is defined by the
wavefunction of the single particle states and product $F_{\boldsymbol{k}%
-q,s^{\prime }}^{\dag }F_{\boldsymbol{k},s}$ defines a band form factor. For
an ideal 2D electron gas (2DEG) we have $F_{\boldsymbol{k},s\sigma }=1$ and
summation over band indices $s,s^{\prime }$ can be ignored. For graphene we
have $F_{\boldsymbol{k},s}=$ $\frac{1}{\sqrt{2}}\left( 
\begin{array}{cc}
1 & se^{i\theta _{\boldsymbol{k}}}%
\end{array}%
\right) ^{\dag }$, where $s=+1$ and $s=-1$ denote conduction and valence
bands, respectively, and $\theta _{\boldsymbol{k}}=\arctan \left(
k_{y}/k_{x}\right) $.

Next, $H$ will be considered as a perturbation Hamiltonian that will cause
transitions from initial state $\left\vert i1,i2\right\rangle =\left\vert
i1\right\rangle \left\vert i2\right\rangle $ with energy $%
E_{i}=E_{1i}+E_{2i} $ to final state $\left\vert f1,f2\right\rangle
=\left\vert f1\right\rangle \left\vert f2\right\rangle $ with energy $%
E_{f}=E_{1f}+E_{2f}$. Here $\left\vert iL\right\rangle $\ ($\left\vert
fL\right\rangle $) is the initial (final) state of layer $L$. The\emph{\ }%
transition rate $\Gamma _{fi}$ from initial state $i$ to final state $f$ is
given by the golden rule formula 
\begin{equation}
\Gamma _{fi}=\frac{2\pi }{\hbar }\left\vert \left\langle f2,f1\right\vert
H\left\vert i1,i2\right\rangle \right\vert ^{2}\delta (E_{i}-E_{f}).
\label{eq:mb_golden_rule}
\end{equation}%
By multiplying $\Gamma _{fi}$ by the energy change $\Delta
E_{1}=E_{1i}-E_{1f}$ and performing an ensemble average over the initial
electronic states, and summing over final electronic states we obtain energy
transfer rate (heat transfer) 
\begin{equation}
P_{12}=\frac{2\pi }{\hbar }\frac{1}{2A}\sum\limits_{\boldsymbol{q}%
}\sum\limits_{i1,f1}\sum\limits_{i2,f2}\Delta E_{1}\widehat{w}_{1i}\widehat{w%
}_{2i}\left\vert M_{\boldsymbol{q}}\right\vert ^{2}\left\vert \left\langle
f1\right\vert \rho _{1\boldsymbol{q}}^{\dag }\left\vert i1\right\rangle
\right\vert ^{2}\left\vert \left\langle f2\right\vert \rho _{2\boldsymbol{q}%
}\left\vert i2\right\rangle \right\vert ^{2}\delta
(E_{1i}+E_{2i}-E_{1f}-E_{2f}),
\end{equation}%
where $\widehat{w}_{Li}$\ is the weighting factor of carrier layer $L$ in
state $i$. We assume that each layer $L$ can be described by a local
temperature $T_{L}$ and, therefore, $\widehat{w}_{Lf}=\widehat{w}_{Li}\exp
[(E_{i}-E_{f})/k_{B}T_{L}]$. \ By using the identity $\delta \left(
E_{A}+E_{B}\right) =\hbar \int_{-\infty }^{+\infty }d\omega \delta \left(
E_{A}-\hbar \omega \right) \delta \left( E_{B}+\hbar \omega \right) $ and
definition of correlator 
\begin{equation}
C_{L}(\boldsymbol{q},\omega )=2\pi \hbar \sum\limits_{n,m}\widehat{w}%
_{Ln}\left\vert \left\langle nL\left\vert \rho _{L\boldsymbol{q}}\right\vert
mL\right\rangle \right\vert ^{2}\delta (E_{Ln}-E_{Lm}+\hbar \omega )
\end{equation}%
we find
\begin{equation}
P_{12}=\frac{1}{2\pi \hbar ^{2}}\left( \frac{1}{2A}\right) ^{2}\int_{-\infty
}^{+\infty }d\omega \sum_{\boldsymbol{q}}\hbar \omega \left\vert M_{%
\boldsymbol{q}}\right\vert ^{2}C_{1}(\boldsymbol{q},-\omega )e^{\hbar \omega
/k_{B}T_{2}}C_{2}(\boldsymbol{q},-\omega ).
\end{equation}%
As we assume internal equilibrium for the different layers we can adopt the
fluctuation-dissipation relation \cite{kubo:1966} $(1-e^{-\hbar \omega
/k_{B}T_{L}})C_{L}(\boldsymbol{q},\omega )=-2\hbar A$Im$\{\chi _{L}(%
\boldsymbol{q},\omega )\}$, where $\chi _{L}(\boldsymbol{q},\omega )$ is the
susceptibility, which can depend on $T_{L}$. Using the
fluctuation-dissipation relation and the property $C_{L}(\boldsymbol{q}%
,-\omega )=e^{-\hbar \omega /k_{B}T_{L}}C_{L}(\boldsymbol{q},\omega )$ we
find the general expression for the interlayer heat transfer 
\begin{equation}
P_{12}=\int_{0}^{+\infty }\frac{d\omega }{2\pi }\sum_{\boldsymbol{q}}\hbar
\omega \left\vert M_{\boldsymbol{q}}\right\vert ^{2}\text{Im}\{\chi _{1}(%
\boldsymbol{q},\omega )\}\text{Im}\{\chi _{2}(\boldsymbol{q},\omega )\}\left[
n_{1}(\hbar \omega )-n_{2}(\hbar \omega )\right] ,  \label{eq:heat_trans}
\end{equation}%
where $n_{L}(\hbar \omega )=\left( \exp (\hbar \omega /k_{B}T_{L})-1\right)
^{-1}$. At the limit $T_{1},T_{2}\rightarrow T$ it is useful to define the
interlayer thermal conductance $G_{12}(T)=P_{12}/(T_{1}-T_{2})$. From Eq. (%
\ref{eq:heat_trans}) we find

\begin{equation}
G_{12}(T)=\frac{1}{4k_{B}T^{2}}\int_{0}^{+\infty }\frac{d\omega }{2\pi }%
\sum_{\boldsymbol{q}}\left( \hbar \omega \right) ^{2}\left\vert M_{%
\boldsymbol{q}}\right\vert ^{2}\frac{\text{Im}\left\{ \chi _{1}(\boldsymbol{q%
},\omega )\right\} \text{Im}\left\{ \chi _{2}(\boldsymbol{q},\omega
)\right\} }{\sinh ^{2}\left( \hbar \omega /2k_{B}T\right) },
\label{eq:thermal_cond}
\end{equation}%
which has a striking similarity with the bilayer drag resistance formula.%
\cite{flensberg:1995b} Equation (\ref{eq:thermal_cond}) has only single
temperature and, therefore, it is more convenient to adopt in the case
studies instead of Eq. (\ref{eq:heat_trans}).

In the following we will assume that the interlayer interaction is mediated
by screened Coulomb interaction, when the matrix element is given by $M_{%
\boldsymbol{q}}=\epsilon _{12}^{-1}(\boldsymbol{q},\omega )U(\boldsymbol{q}%
)F_{12}(d),$ where $U(\boldsymbol{q})=e^{2}/2\varepsilon _{b}q$ is the 2D
Fourier transform of Coulomb potential ($\varepsilon _{b}$ is the background
dielectric constant) and $F_{12}(d)$ is the spatial form factor, which
depends on the spatial extent of the electron wave functions and layer
distance $d$. For graphene the extend is practically zero and for the sake
of simplicity here we assume vanishing extent for the semiconductor systems
as well. Thus, we use $F_{12}(d)=\exp (-qd)$. The inter-layer dielectric
function $\epsilon _{12}(\boldsymbol{q},\omega )$ is given by $\epsilon
_{12}(\boldsymbol{q},\omega )=\left[ 1-U(\boldsymbol{q})\chi _{1}(%
\boldsymbol{q},\omega )\right] \left[ 1-U(\boldsymbol{q})\chi _{2}(%
\boldsymbol{q},\omega )\right] -F_{12}^{2}U(\boldsymbol{q})^{2}\chi _{1}(%
\boldsymbol{q},\omega )\chi _{2}(\boldsymbol{q},\omega )$.\cite%
{rojo:1999,flensberg:1995b}

In the ballistic limit the carrier mean free path $l_{e}$ exceeds the layer
distance ($l_{e}\gg d$ ) and we use the ideal 2D susceptibilities. For 2DEG
we have \cite{stern:1967} 
\begin{equation}
\chi _{L}(\boldsymbol{q},\omega )=\nu (2z)^{-1}\left[ 2z-\Omega
_{-}(z,u)-\Omega _{+}(z,u)+\digamma _{-}(z,u)-\digamma _{+}(z,u)\right] ,
\label{eq:X0_ball}
\end{equation}%
where $\Omega _{\pm }(z,u)=C_{\pm }\sqrt{\left( z\pm u\right) ^{2}-1}$, $%
\digamma _{\pm }(z,u)=iD_{\pm }\sqrt{1-\left( z\pm u\right) ^{2}}$, $%
z=q/2k_{F}$, $u=\omega /qv_{F}$, $C_{\pm }=\left( z\pm u\right) /\left\vert
z\pm u\right\vert \ $and $D_{\pm }=0$ for $\left\vert z\pm u\right\vert >1$,
and $C_{\pm }=0\ $and $D_{\pm }=1$ for $\left\vert z\pm u\right\vert <1$.
Here $v_{F}$ ($k_{F}$) is the Fermi velocity (wave vector) and $\nu =\nu
(\varepsilon _{F})$ is the density of states at Fermi level $\varepsilon
_{F}\gg k_{B}T$. For ballistic graphene the expression for $\chi _{L}(%
\boldsymbol{q},\omega )$ is quite lengthy and will not be presented here. It
can be found, for example, from Ref. \cite{hwang:2007}.

\subsection{Approximation formulas}

Even though there are some fundamental differences between graphene and
2DEGs, the response of these systems is similar at low frequencies and small 
$q$. Indeed, for the Taylor series expansion of 2DEG [Eq. (\ref{eq:X0_ball}%
)] and graphene susceptibilities \cite{hwang:2007} we find the same result 
\begin{equation}
\chi _{L}(\boldsymbol{q},\omega )\simeq -\nu (1+i\frac{\omega }{v_{F}q}).
\label{eq:X0_ball_appr}
\end{equation}%
Respectively, in the diffusive limit ($\omega \tau ,l_{e}/d\ll 1$) the
susceptibility can be approximated as 
\begin{equation}
\chi _{L}(\boldsymbol{q},\omega )\simeq -\nu \frac{iDq^{2}}{\omega +iDq^{2}},
\label{eq:X0_diff}
\end{equation}%
where $D=v_{F}^{2}\tau /2$ is the diffusion coefficient and $\tau
=l_{e}/v_{F}$ is the momentum relaxation time.

By using Eq. (\ref{eq:X0_ball_appr}) in Eq. (\ref{eq:thermal_cond}) for two
similar ballistic 2DEG and graphene layers we find asymptotic
low-temperature result 
\begin{equation}
G_{12}(T)\simeq \frac{f\left( \kappa d\right) }{2d^{2-\alpha }}\frac{\hbar }{%
v_{F}^{\alpha }}\left( \frac{k_{B}}{\hbar }\right) ^{2+\alpha }T^{1+\alpha },
\label{eq:G_ball}
\end{equation}%
where $\alpha =1.9$, $\kappa =\frac{\nu e^{2}}{2\varepsilon _{b}}$ is the
screening wave vector and $f\left( a\right) \simeq 
(a^{-2}+2.21a+1.24)^{-1}$. The above Equation provides a good approximation
when $k_{F}dk_{B}T<2E_{F}$. Note that parameter $\kappa d$ characterizes the
screening of the interlayer interaction: large (small) $\kappa d$ means
strong (weak) screening.

In the diffusive limit we use Eq. (\ref{eq:X0_diff}) in Eq. (\ref%
{eq:thermal_cond}) and we find low temperature approximation formula 
\begin{equation}
G_{12}(T)\simeq \dfrac{3A_{3}}{16\pi }\dfrac{\varepsilon _{b}}{d}\dfrac{1}{%
2\sigma }\hbar \left( \dfrac{k_{B}}{\hbar }\right) ^{3}T^{2}.
\label{eq:G_diff}
\end{equation}%
Here $\sigma =e^{2}\nu D$ is the DC conductivity of single layer and $%
A_{n}=\Gamma (n)\zeta (n)=\int dxx^{n-1}/\left[ \exp (x)-1\right] $.
Equation (\ref{eq:G_diff}) is applicable when $(l_{e}/v_{F})k_{B}T/\hbar
<\left( l_{e}/d\right) ^{2}$. Note that the diffusive $G_{12}$ [Eq. (\ref%
{eq:G_diff})] greatly exceeds the one in the ballistic case [Eq. (\ref%
{eq:G_ball})], which is the manifestation of enhanced fluctuations and
dissipation due to disorder.

\subsection{Electron-phonon coupling}

As depicted in the thermal circuit of Fig. \ref{fig:schematic}(b) $G_{12}$
competes with the electron-phonon thermal conductance $G_{Lp}$, which at the
limit $T_{L},T_{p}\rightarrow T$ is given by $G_{Lp}(T)=P_{Lp}/\left(
T_{L}-T_{p}\right) $. In 2DEGs at low temperatures the electron-phonon
energy transfer is dominated by screened deformation potential ($G_{Lp}^{DP}$%
) and piezoelectric ($G_{Lp}^{PE}$) interaction with total thermal
conductance $G_{Lp}(T)=G_{Lp}^{DP}(T)+G_{Lp}^{PE}(T)$. For the deformation
potential contribution we have \cite{price:1982,prunnila:2007} 
\begin{equation}
G_{Lp}^{DP}(T)=\sum_{\lambda }\frac{F_{6-n}2^{n}}{l_{e}^{n}\kappa ^{2}}%
\left\langle f_{n}(\theta )\Xi ^{2}\right\rangle v_{\lambda }^{n-6}T^{6-n},
\label{eq:G_ep_DP}
\end{equation}%
where $n=0(1)$ represents ballistic (diffusive) limit of electron-phonon
coupling, for which we have $q_{\lambda T}l_{e}>1(<1)$. Here $q_{\lambda
T}=k_{B}T/\hbar v_{\lambda }$ is the thermal phonon wave vector, factor $%
F_{i}=\frac{\nu A_{i}}{2\pi ^{2}\rho v_{F}}\frac{(i+1)k_{B}^{i+1}}{\hbar ^{i}%
}$, $v_{L(T)}$ is the longitudinal (transversal) phonon velocity and $\rho $
is the mass density of the crystal. The brackets $\left\langle \cdots
\right\rangle $ stand for solid angle average and $\theta $ is the angle
with respect to the z-axis, which is perpendicular to the layer\ plane. $%
\left\langle f_{n}(\theta )\Xi ^{2}\right\rangle $ is an effective
deformation potential coupling and $f_{0}(\theta )=\sin \theta $ and $%
f_{1}(\theta )=\frac{\sin ^{2}\theta }{\alpha +\sin ^{2}\theta }$. Parameter 
$\alpha =(\kappa l_{e}v_{F}/v_{\lambda })^{-2}$ and as a result $%
f_{1}(\theta )$ $\approx 1$. The piezoelectric coupling gives rise to
contribution \cite{price:1982,prunnila:2007,khveshchenko:1997} 
\begin{equation}
G_{Lp}^{PE}(T)=\sum_{\lambda }\frac{F_{4-n}2^{n}}{l_{e}^{n}\kappa ^{2}}%
\left\langle f_{n}(\theta )K^{2}\right\rangle v_{\lambda }^{n-4}T^{4-n},
\label{eq:G_ep_piezo}
\end{equation}%
where $\left\langle f_{n}(\theta )K^{2}\right\rangle $\ is the effective
piezo coupling. For graphene the electron-phonon coupling is dominated by
deformation potential coupling \cite{viljas:2010} and vector potential
coupling \cite{chen:2012}\ ($G_{Lp}^{VP}(T)$) giving $%
G_{Lp}(T)=G_{Lp}^{DP}(T)+G_{Lp}^{VP}(T)$. For these both contributions we
will use directly the results of Ref. \cite{chen:2012}.

\section{Results and discussion}

In this section, we calculate the interlayer thermal conductance $G_{12}$ of
selected semiconductor and graphene bilayer systems at the ballistic and
diffusive limit and discuss possible experimental configurations to
investigate $G_{12}$. Interlayer thermal conductance will be compared to
electron-phonon thermal conductance $G_{Lp}$. Diffusive $G_{12}$ is
considered at larger interlayer separation than the ballistic one in order
to assure that the diffusive response formula [Eq. (\ref{eq:X0_diff})] is
valid and condition $k_{F}l_{e}>1$ is fulfilled.

\subsection{Calculations for different bilayers}

\begin{figure}[b]
\centering\includegraphics[width=0.53\textwidth]{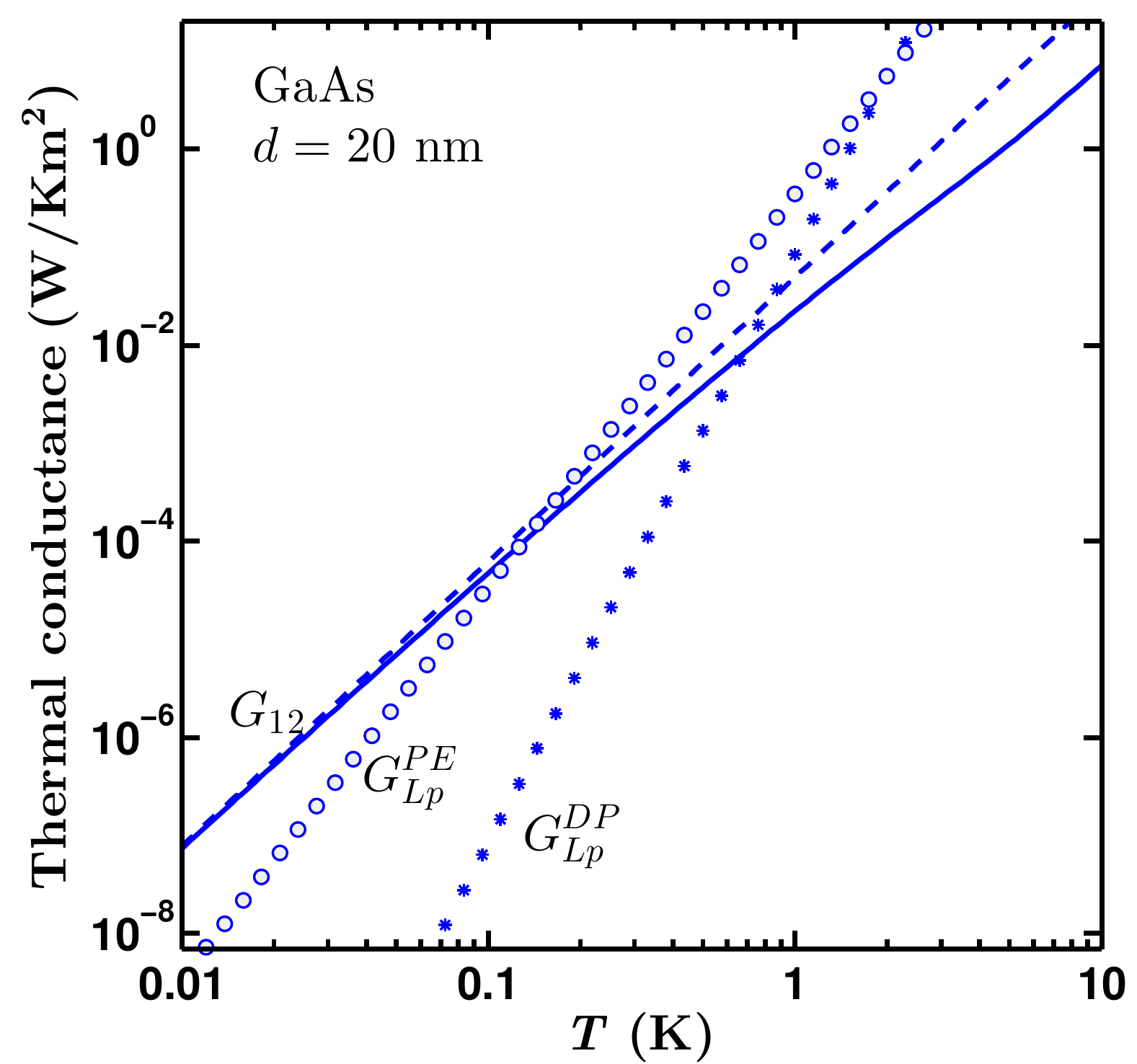}
\caption{(color online) The interlayer thermal conductance ($G_{12}$) and
deformation potential ($G_{Lp}^{DP}$) and piezoelectric ($G_{Lp}^{PE}$)\
electron-phonon thermal conductance at the ballistic limit for GaAs bilayer
system with electron density $n=10^{15}$ m$^{-2}$. Solid curve is the result
from numerical integration using the ballistic response function. Dashed
curve obtained using Eq. (\protect\ref{eq:G_ball}).}
\label{fig:ballistic}
\end{figure}
Figure \ref{fig:ballistic} shows $G_{12}(T)$ obtained numerically from Eqs. (%
\ref{eq:thermal_cond}) and (\ref{eq:X0_ball}) in the case of symmetric high
mobility (ballistic) GaAs bilayer\cite{gramila:1991} [depicted in Fig. \ref%
{fig:schematic}(c)] with single layer electron density $n=10^{15}$ m$^{-2}$
and $d=20$ nm. Asymptotic limit formula of Eq. (\ref{eq:G_ball}) is also
plotted. In the phonon contribution we have $\sum_{\lambda }\left\langle
\sin \theta \Xi ^{2}\right\rangle v_{\lambda }^{-6}=\frac{1}{4}\pi \Xi
_{d}^{2}v_{L}^{-6}$, where $\Xi _{d}=10$ eV is the dilatational deformation
potential constant, and $\sum_{\lambda }\left\langle \sin \theta
K^{2}\right\rangle $ $v_{\lambda }^{-4}=\left( \frac{e}{\varepsilon _{b}}%
\right) ^{2}e_{14}^{2}\allowbreak \pi \left( \frac{89}{1024}v_{L}^{-4}+\frac{%
107}{1024}v_{T}^{-4}\right) $, where $e_{14}=-0.16$ C/N is the only non-zero
element of the piezotensor. Other parameters can be found from Ref. \cite%
{madelung:2004}. Equations (\ref{eq:G_ep_DP}) and (\ref{eq:G_ep_piezo}) are
plotted as symbols in Fig. \ref{fig:ballistic} in the ballistic limit of
electron-phonon coupling. Below few Kelvin piezoelectric coupling fully
dominates and as a result the temperature regime where $G_{Lp}<G_{12}$ is
pushed towards relatively low, but experimentally achievable, temperatures.
The crossover occurs at $T\sim 140$ mK.

\begin{figure}[htbp]
\centering\includegraphics[width=0.53\textwidth]{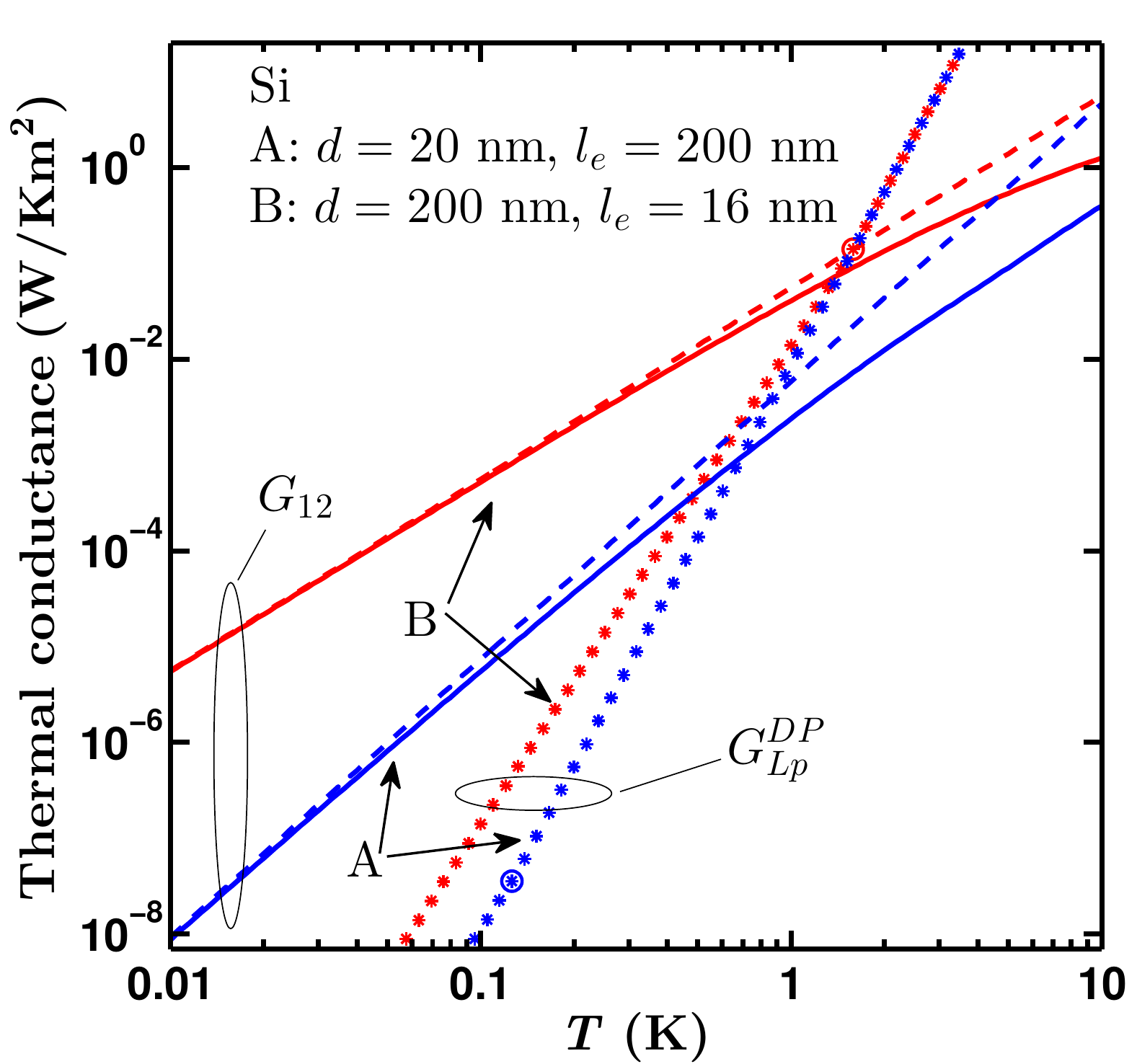}
\caption{(color online) The interlayer thermal conductance ($G_{12}$) and
deformation potential ($G_{Lp}^{DP}$)\ electron-phonon thermal conductance
for two Si bilayer devices with $n=5\times 10^{15}$ m$^{-2}$. For device A
(B) solid curves are results from numerical integration using the ballistic
(diffusive) susceptibility and dashed curves are obtained from the
approximation formula of Eq. (\protect\ref{eq:G_ball}) (Eq. (\protect\ref%
{eq:G_diff})). The circle marks the crossover where electron-phonon coupling
changes from ballistic to diffusive.}
\label{fig:diffusive}
\end{figure}
For silicon based bilayer [see Ref. \cite{takashina:2004,*prunnila:2005} and
Fig. \ref{fig:schematic}(d)] we consider both ballistic and diffusive limits
at electron density $n=5\times 10^{15}$ m$^{-2}$. Parameters for Si can be
found from Ref. \cite{madelung:2004}. The curves in Fig. \ref{fig:diffusive}
are calculated for symmetric high (low) mobility Si bilayer system with
mobility $\mu =2.5$ ($0.2$) m$^{2}/$Vs, mean-free path $l_{e}=200$ ($16$) nm
and layer distance $d=20$ ($200$) nm. For the high mobility device we have
used the ballistic limit response [Eq. (\ref{eq:X0_ball})] and for the low
mobility one the diffusive response [Eq. (\ref{eq:X0_diff})]. Silicon is not
piezoelectric so for $G_{Lp}$\ we need to consider only $G_{Lp}^{DP}$.Due
finite electron mean free path (even for the high mobility device) we will
include ballistic and diffusive limits of $G_{Lp}^{DP}$. For simplicity we
plot $G_{Lp}^{DP}$ so that it changes abruptly from diffusive to ballistic
formula [note that Eq. (\ref{eq:G_ep_DP}) is not valid close to $q_{\lambda
T}l_{e}=1$]. For Si 2DEG we have $\sum_{\lambda }\left\langle \sin \theta
\Xi ^{2}\right\rangle v_{\lambda }^{-6}=\frac{1}{32}\pi \left( 4\Xi _{d}\Xi
_{u}+8\Xi _{d}^{2}+\Xi _{u}^{2}\right) v_{L}^{-6}+\frac{1}{32}\pi \Xi
_{u}^{2}v_{T}^{-6}$ and $\sum_{\lambda }\left\langle \Xi ^{2}\right\rangle
v_{\lambda }^{-6}=\left( \frac{2}{3}\Xi _{d}\Xi _{u}+\Xi _{d}^{2}+\frac{1}{5}%
\Xi _{u}^{2}\right) v_{L}^{-6}+\frac{2}{15}\Xi _{u}^{2}v_{L}^{-6}$, where $%
\Xi _{u}$ is the uniaxial deformation potential constant. We use the typical
values $\Xi _{d(u)}=-11.7(9.0)$ eV. For the high and low mobility Si systems
the crossover temperature where $G_{12}=G_{Lp}$ is $660$ mK and $1.4$ K,
respectively. Even though we have set $d$ an order of magnitude larger for
the diffusive device, still the crossover occurs at higher temperature,
which is the signature of the enhancement of fluctuations/dissipation and,
thereby, interlayer coupling due to disorder. Note that deformation
potential electron-phonon coupling is also enhanced due to disorder.

Next we consider graphene bilayer\cite{kim:2011} that is depicted in Fig. %
\ref{fig:schematic}(e). 
\begin{figure}[hbp]
\centering\includegraphics[width=0.53\textwidth]{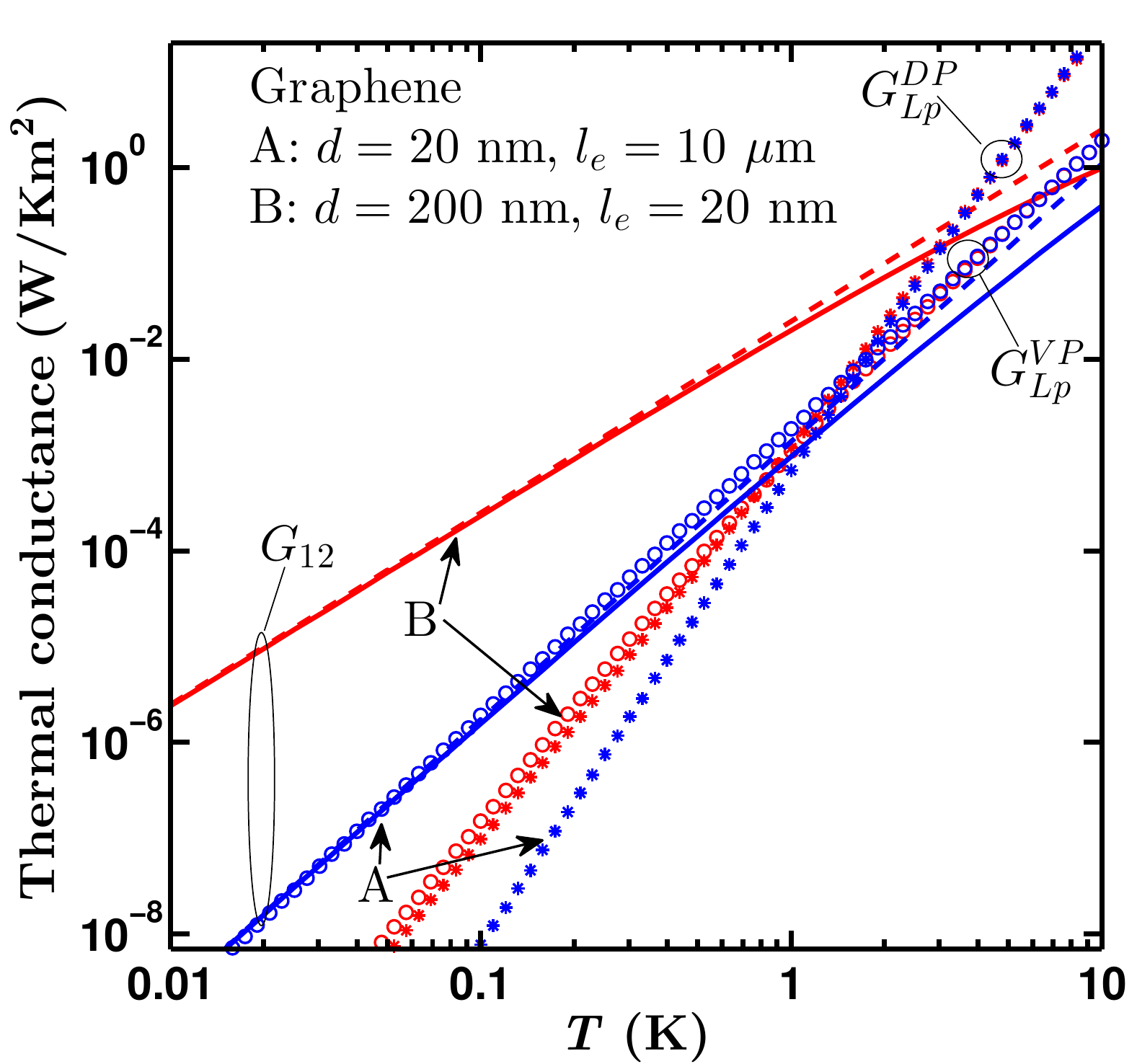}
\caption{(color online) The interlayer thermal conductance ($G_{12}$),
deformation potential ($G_{Lp}^{DP}$)\ and vector potential ($G_{Lp}^{VP}$)
electron-phonon thermal conductance for two graphene bilayer devices with $%
n=1\times 10^{16}$ m$^{-2}$. $G_{Lp}^{DP}$ and $G_{Lp}^{VP}$\ are from Ref. 
\protect\cite{chen:2012}. For device A (B) solid curves are results from
numerical integration using the ballistic (diffusive) susceptibility and
dashed curves are obtained from the approximation formula of Eq. (\protect
\ref{eq:G_ball}) (Eq. (\protect\ref{eq:G_diff})).}
\label{fig:graphene}
\end{figure}
The curves in Fig. \ref{fig:graphene} are calculated for symmetric high
(low) mobility device with mobility $\mu =1.59$ ($0.17$) m$^{2}/$Vs,
mean-free path $l_{e}=10$ $\mu $m ($20$ nm), layer distance $d=20$ ($200$)
nm and electron density $n=10\times 10^{15}$ m$^{-2}$. We have used $%
v_{F}=10^{6}$\ m/s and assumed AlO dielectric between the layers. \ As
above, for the high mobility and for low mobility device we use ballistic
and diffusive response functions, respectively. Screened deformation
potential and vector potential electron-phonon contributions are plotted in
Fig. \ref{fig:graphene} as symbols. For high mobility graphene $G_{Lp}^{VP}$
dominates at the lowest temperatures and the cross-over where $G_{12}=G_{Lp}$
occurs at relatively low temperature of $T=40$ mK. The disorder enhancement
of the interlayer heat transfer pushes the threshold for low mobility
graphene to $T=3.0$ K. Note that the vector potential electron-phonon
coupling is decreased with the disorder in contrast to deformation potential
coupling.

Another widely explored semiconductor bilayer carrier system, which can be
realized using compound semiconductors\cite{sivan:1992} or Si\cite%
{prunnila:2008,takashina:2009}, is the electron-hole bilayer. The complexity
of the valence band makes this system more difficult to analyze
theoretically. We will not present $G_{12}$ for such system here explicitly,
but it should behave in similar fashion as its electron-electron counter
part. However, one thing that may differ drastically from electron-electron
bilayer system is the carrier-phonon coupling. Due to asymmetry in the
deformation potential coupling between the different layers the
carrier-phonon coupling can be unscreened and as a result $G_{Lp}$ can be
strongly enhanced at low temperatures. \cite{prunnila:2007} The enhancement
factor depends on the details of the system, but in many cases it is of the
order of $\left( \kappa /q_{\lambda T}\right) ^{2}$, which suggests that for
semiconductor electron-hole bilayers $G_{Lp}$ can dominate over $G_{12}$
even down to very low temperatures. Note that also in symmetric electron
bilayers $G_{Lp}$ can be affected by the presence of another carrier system
in non-trivial way, but significant enhancement is not expected\cite%
{prunnila:2007}.

\subsection{Possible experimental realizations}

The interlayer heat transfer can be investigated experimentally by varying
the input powers $P_{L}$ while measuring the electron temperatures $T_{L}$
[see Fig. \ref{fig:schematic}(b)]. Uniform input power follows, for example,
from Ohmic heating. This technique has been broadly utilized in the
electron-phonon coupling measurements. Indeed, the electron-phonon
contributions $G_{Lp}$ can be investigated independently from $G_{12}$ at
balanced input power that gives $T_{1}=T_{2}$. In the case of semiconductor
bilayer the other layer can be also depleted to get a handle on $G_{Lp}$.
Note that the Ohmic heating technique has been utilized also in the
investigation of coupling of Johnson-Nyquist noise heating between two
resistors at different temperature\cite{meschke:2006}, which is conceptually
very close to the case presented here.

It is not necessarily trivial to measure electron temperature of the
individual layers. For example, quantum corrections of resistivity and
Shubnikov- de Haas oscillations, that have been used as electron
thermometer, can be affected by the other layer in a complicated way. More
local temperature probes based, e.g., on quantum point contacts and quantum
dots have also been investigated.\cite%
{appleyard:1998,prance:2009,gasparinetti:2012} Noise thermometry provides an
attractive way to probe electron temperature. It has been recently used for
single layer graphene \cite{fong:2012} and could be adopted in
investigations of $G_{12}$.

Interlayer heat transfer can also be investigated in more indirect way by
coupling the individual layers to metallic electrodes, to which the input
power is fed and where temperature is sensed. As metals have quite large
electron-phonon coupling the volume of the metallic islands should be
sufficiently small in order not to hide $G_{12}$. Especially in the case of
Si, doped contact regions can serve as the metallic islands. This is
attractive approach as the electron-phonon coupling in doped semiconductors
can be relative weak so that $G_{12}$ still dominates. The sign of $P_{L}$
can be also reversed, which is equivalent to cooling. This can be achieved
by quantum dots \cite{prance:2009} or semiconductor-superconductor contacts 
\cite{savin:2001}.

As $G_{12}$ (and $G_{Lp}$) depends on the electron densities and on the
interlayer density balance it is desirable to adjust the layer densities by
external gates. In general, $G_{12}$ could be used as a gate voltage
controlled thermalization path (thermal switch). However, it is important to
note (as already pointed out in Ref.\cite{prunnila:2007}) that similar
near-field thermal coupling to $G_{12}$ can exist between the 2D carriers
and the external gate electrodes.

\section{Summary and conclusions}

In summary, a near-field heat transfer effect due to interlayer interaction
in bilayer carrier systems was investigated. By using perturbation theory
and fluctuation-dissipation relations we derived a general expression of
near-field interlayer energy transfer rate [Eq. (\ref{eq:heat_trans})] and
thermal conductance [Eq. (\ref{eq:thermal_cond})]. Our formulation can be
applied, for example, to semiconductor and graphene based bilayers. We
presented analytical approximation formulas and detailed calculations{\large %
\ }of interlayer heat transfer due to screened Coulomb interaction for GaAs,
Si and graphene based bilayers. It was shown that remarkably the interlayer
heat transfer can dominate over the electron-phonon coupling to the thermal
bath below a crossover temperature that depends on the system parameters. We
found a crossover temperature of $140$ mK ($660$ mK) for ballistic GaAs (Si)
bilayer with $d=20$ nm layer distance and carrier density $n=10^{15}$ m$%
^{-2} $ ($5\times 10^{15}$ m$^{-2}$). Strong vector potential
electron-phonon coupling in ballistic graphene results in low crossover
temperature of $40$ mK ($n=10\times 10^{15}$ m$^{-2}$). Interlayer heat
transfer is enhanced by disorder and for low mobility Si (graphene) bilayer
with $d=200$ nm the crossover occurs already at $\sim 1.4$ K ($3.0$ K). The
crossover temperatures reported here can be accessed by standard
experimental equipment and we introduced possible experimental
configurations to investigate the interlayer heat transfer.

Finally, we note that by lowering the electron densities and/or by
increasing the temperature plasmons and virtual phonons may start to play a
role in the interlayer interaction. These excitations are known to enhance
the bilayer drag effect \cite{flensberg:1994,bonsager:1998} and the
enhancement should also be observable in the interlayer heat transfer. In a
very dilute and strongly interacting systems enhancement of drag, that
cannot be explained with plasmons or virtual phonons, has also been observed.%
\cite{pillarisetty:2002} Therefore, depending on the system parameters the
crossover temperature, below which the interlayer heat transfer starts to
dominate over the electron-phonon coupling to the thermal bath, can be
significantly higher than the ones given in this work. Studies of plasmonic
effects, virtual phonon excitations, dilute carrier regime and elevated
temperatures are left for future investigations. At elevated temperatures
the effect described in this paper can also be of relevance for inter-flake
heat transfer in thermal interface materials fabricated from graphene
composites.\cite{Shahil:2012} The concepts presented in this work can be
extended to coupled one-dimensional carrier systems.

\begin{acknowledgments}
The authors want to acknowledge useful discussions with P.-O. Chapuis, K.
Flensberg and D. Gunnarsson. This work has been partially funded by the
Academy of Finland through grant 252598 and by EU through project 256959
NANOPOWER.
\end{acknowledgments}

\bibliographystyle{apsrev4-1}
\bibliography{2D_2D_heat_refs}


\end{document}